\documentclass{article}

\PassOptionsToPackage{numbers, compress, square}{natbib}
\usepackage[preprint]{neurips_2022}

\usepackage[utf8]{inputenc} 
\usepackage[T1]{fontenc}    
\usepackage{hyperref}       
\usepackage{url}            
\usepackage{booktabs}       
\usepackage{amsfonts}       
\usepackage{nicefrac}       
\usepackage{microtype}      
\usepackage{xcolor}         
\usepackage{float}          
\usepackage{graphicx}       
\usepackage{caption}        
\usepackage{amsmath}        

\usepackage{braket}
\usepackage{microtype}

\DeclareMathOperator{\tr}{tr}
\newcommand{\ketbra}[2]{{\left|{#1}\right\rangle\!\left\langle{#2}\right|}}

\usepackage{array}
\newcolumntype{P}[1]{>{\centering\arraybackslash}p{#1}}
\usepackage{algorithm}
\usepackage{algpseudocode}

\usepackage{xcolor}
\newcommand{\nocontentsline}[3]{}
\newcommand{\tocless}[2]{\bgroup\let\addcontentsline=\nocontentsline#1{#2}\egroup}

\title{Image Classification by Throwing Quantum Kitchen Sinks at Tensor Networks}
\author{%
  Nathan X.~Kodama\\
  Case Western Reserve University\\
  Cleveland, OH
  \And
  Alex Bocharov\\
  Microsoft Quantum\\
  Redmond, WA\\
  \And
  Marcus P. da Silva\\
  Microsoft Quantum\\
  Redmond, WA
}

\definecolor{lightblue}{HTML}{4772B8}
\definecolor{darkred}{HTML}{b84747}
\definecolor{gold}{HTML}{e5a600}
\definecolor{grey}{HTML}{999999}

\begin{document}
\normalsize
\maketitle

\begin{abstract}
Several variational quantum circuit approaches to machine learning
have been proposed in recent years, with one promising class of
variational algorithms involving tensor networks operating on states
resulting from \textit{local} feature maps. In contrast, a random
feature approach known as {\em quantum kitchen sinks} provides
comparable performance, but leverages \textit{non-local} feature
maps. Here we combine these two approaches by proposing a new circuit
ansatz where a tree tensor network coherently processes the
non-local feature maps of quantum kitchen sinks, and we run numerical
experiments to empirically evaluate the performance of the new ansatz 
on image classification.  From the perspective of
classification performance, we find that simply combining quantum
kitchen sinks with tensor networks yields no qualitative
improvements. However, the addition of {\em feature optimization}
greatly boosts performance, leading to state-of-the-art 
quantum circuits for image classification, requiring only shallow
circuits and a small number of qubits -- both well within
reach of near-term quantum devices.
\end{abstract}

\section{Introduction}
\label{sec:introduction}
Tensor network (TN) methods have been studied across physics,
mathematics, and computer science for their expressive power,
interpretability, and computational
efficiency~\cite{Kolda2009,Eisert2013,Vervliet2014,Cichocki2015,Sidiropoulos2016,Orus2019,Biamonte2019}. These
properties make them well-suited for machine learning, recently
leading to several promising results~\cite{Novikov2015,
  Stoudenmire2016, Cohen2016, Glasser2019}. Under some mild
constraints, a TN may be interpreted as a quantum
circuit~\cite{Grant2018,Huggins2019,Haghshenas2022}. In some cases,
these circuits may be simulated with a computational effort that
scales polynomially with the number of qubits, which makes them
amenable to numerical exploration on classical computers---in stark
contrast with the exponential scaling expected in simulation of
arbitrary quantum circuits.

One of the underlying challenges in developing quantum TN methods for
machine learning is to establish a competitive performance
baseline. A classical algorithm, random kitchen sinks 
(RKS)~\cite{Rahimi07,Rahimi08a,Rahimi08b}, offers some inspiration since 
it has been shown to be competitive with multilayer neural 
networks~\cite{May2017}. The principal difference between the two approaches
is that RKS, a kernel method, replaces the costly learning of low-level
features in a multilayer network with random non-local features. This 
randomization step may be considered a special kind of feature engineering 
in a machine learning pipeline. 

A quantum algorithm based on RKS, known as {\em quantum kitchen sinks} 
(QKS)~\cite{Wilson2018}, has been shown to produce error rates that are comparable 
to quantum TN methods~\cite{Huggins2019} at similar qubit utilizations and 
minimal classical overhead. An important question is whether coherent quantum 
processing can provide any improvement in classification performance. The
original QKS results indicated some mild improvement with coherent
processing over a small number of qubits and some degradation at a
large number of qubits---and no systematic guidance for how coherent
processing could be leveraged. Here we attempt to shed some light into
this question by combining QKS and TN circuits
into a new variational ansatz, and show this combined ansatz compares
favorably to both of its ancestors.

In both the classical and quantum contexts, the contest between the
multilayer networks and relatively shallow (Q/R)KS models is strongly 
impacted by the design of the network layers and 
features. It is well-known that any function can be learned by either shallow 
or multilayer architecture; the essential questions are (1) how efficient is 
the corresponding architecture and (2) how much work is required
in optimizing the architecture to the task~\cite{Bengio2007}. We
assess each of these points in turn by separately studying the two
modules of the combined QKS and TN protocol.

\begin{figure}[h!]
  \centering
  \includegraphics[width=\textwidth]{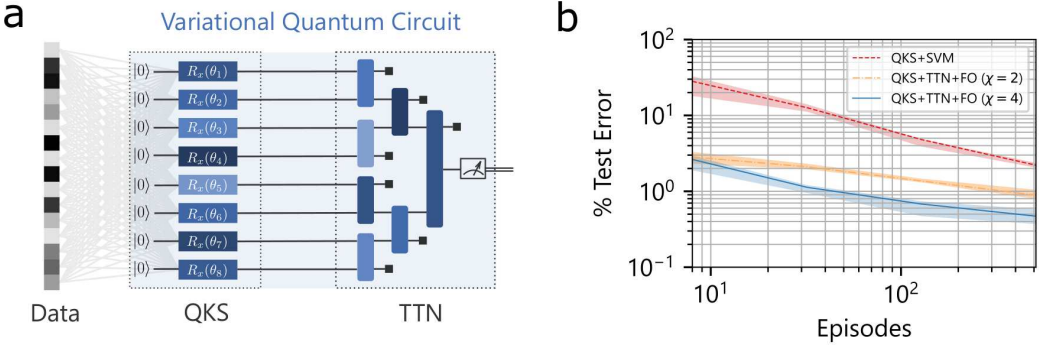}
  \caption{(a) Classical data parameterize rotations in quantum kitchen sinks (QKS).
  Application of quantum rotation gates such as $R_P(\theta)=\cos(\theta/2)I - i\,\sin(\theta/2)P$ (where $P$ is a Pauli operator) results in quantum states whose amplitudes are mixtures of non-linear functions of the classical angles.
   Variational circuits where such quantum gates are followed with a tree tensor network (TTN) structure coherently process the output of QKS before a small part of the state is measured.
   (b) Test errors over 10 realizations (median and 68\% credibility intervals) as a function of the number of episodes. Joint optimization of the rotations and TTN lead to binary classifiers with error rates below $1\%$ (on ``3'' vs. ``5'' handwritten digits from the MNIST dataset).}
  \label{fig:fig1}
\end{figure}

A concrete example of a TN is a {\em tree tensor network} (TTN),
depicted in (\hyperref[fig:fig1]{Fig. 1a}). In these networks, two
collections of $n$ qubits each (where $\chi=2^n$ is referred to as the
{\em bond dimension}) interact unitarily, but only one of the
resulting collections of $n$ qubits continues on for additional
computation. We refer to the quantum circuit that results from
combining QKS with TTN as QKS+TTN, in contrast to the previous QKS
approach which leveraged linear classifiers (such as support vector
machines\footnote{The work of~\citet{Wilson2018} used logistic
regression, while we use linear support vector machines here for
convenience -- we do not expect a material performance difference
between the two.}), which we refer to as QKS+SVM. In this work, we only
considered $\chi=2$ and $\chi=4$ and focused on the TTN to build a
large coherent computation. An attractive feature of the proposed TTN
and other related ansatze (such as multi-scale entanglement
renormalization ansatz~\cite{Evenbly2009}) is that they
avoid the problem of {\em barren optimization landscapes}, due to the
shallow circuit
depth~\cite{McClean2018,Wang2020,Cerezo2021,Arrasmith2021},
and their sparse connectivity prevents errors from accumulating
pathologically~\cite{Kim2017,Kim2019,Anikeeva2021}, making the
implementation of these circuits on near-term hardware appealing.

Classification using QKS+TTN requires simulating the evolution of the quantum state through fixed circuits (parameterized by the classical input data), which are called kitchen sinks or \textit{episodes}~\cite{Wilson2018}, and the TTN, which concludes with the application of a measurement on the remaining $n$ qubits at the root of the tree~\cite{Grant2018,Huggins2019}. For a {\em single-shot} execution of the circuit, the resulting outcome corresponds to the classical output of the classifier. For a {\em multi-shot} execution of the circuit, we consider the application of a linear classifier to the observed outcome frequencies. For binary classification, we may consider measuring a single qubit out of the remaining $n$, but more general classification tasks may use all possible $\chi$ outcomes of the final measurement.

\textbf{Contributions}. We propose a new circuit ansatz where a TTN coherently processes the non-local feature maps of QKS. We empirically evaluate the performance of the new ansatz on image classification. In terms of classification performance, we find that simply combining QKS and TTN yields no qualitative improvements. However, training QKS+TTN with \textit{feature optimization} (QKS+TTN+FO) significantly boosts performance and improves qubit utilization over the QKS+SVM baseline (\hyperref[fig:fig1]{Fig. 1b}), leading to state-of-the-art quantum circuits for image classification, while requiring only shallow quantum circuits and a small number of qubits -- both well within reach of near-term quantum devices.

A crucial feature of the new ansatz is how classical data are mapped to quantum states. Previous work considered \textit{local feature maps}~\cite{Grant2018,Huggins2019}, where each classical dimension $x_i$ was encoded in one qubit amplitude $\cos(x_i)\ket{0}+\sin(x_i)\ket{1}$ (or similar non-linear encoding), which required a number of qubits equal to the number of classical features in the problem (synthetic features may be naturally included as desired). Our work uses the \textit{non-local feature map} provided by QKS so that the quantum state corresponding to each episode contains information about the entire data. This decouples the correlation structure in the data from the quantum circuit structure so that we may apply the same TN structure to datasets of different dimensionality and correlation structure.

\section{Methods and Benchmarks}
\label{sec:methods}
We begin by describing the datasets and the methods used for QKS+SVM, QKS+TTN, and QKS+TTN+FO. We present two approaches to reducing the complexity of the overall model: (\textit{i}) translational symmetry in the TTN and (\textit{ii}) sparsity in the features.

\textit{Benchmarks}. The MNIST dataset~\cite{LeCun1998} is a well-known benchmark in machine learning. Each digit is a $(28 \times 28)$-pixel, 8-bit grayscale image of a handwritten digit. We treat each image as a vector with dimension $28^2=784$ and split the dataset into 60,000 training and 10,000 testing images. While it is a benchmark for multi-class classification, we choose to focus on the binary classification two digits that are difficult to distinguish: the handwritten digits ``3'' and ``5'', which we refer to as \textit{(3,5)-MNIST}. All licensing information for existing assets can be found in the supplementary material.

\textbf{Establishing a baseline with quantum kitchen sinks}. Consider an unknown target function $f: {\tt Data} \rightarrow {\tt Labels}$. The general approach to ``learning'' the target function is to approximate it with some structured parameterized function
$g_{\boldsymbol\theta}$ taken from some \emph{hypothesis space}. In deep learning, such  $g_{\boldsymbol\theta}$ is structured as a parameterized composition of a large number of simple non-linear functions, and the many parameters of this composition all have to be learned for optimal performance. As an alternative, Rahimi and Recht~\cite{Rahimi07,Rahimi08a,Rahimi08b} have shown that it was possible to produce a sufficiently rich set of hypothesis functions $g(\cdot,\boldsymbol\theta)$ as weighted linear sums of simple non-linear functions with \emph{random} parameters. The weights in the summary linear combination of such non-linear terms still need to be learned, but this is a linear learning step that can be done at a significantly lower cost than learning a full customary neural network. Results presented in~\cite{Rahimi07,Rahimi08a,Rahimi08b} suggest that there are multiple reasonable choices of non-linearities that can be used as structural blocks for RKS: cosine, sign, and indicator functions in particular.

\begin{figure}[h!]
  \begin{center}
  \includegraphics[scale=0.75]{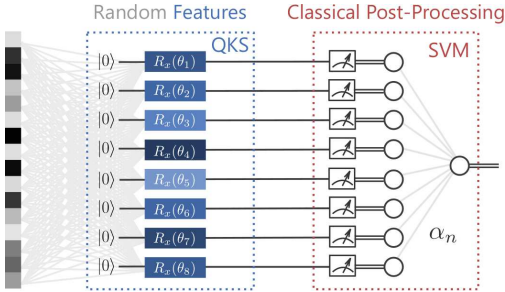}
  \end{center}
  \caption{A quantum kitchen sink (QKS), highlighted in 
  \textcolor{lightblue}{blue}, is made up of
    independent episodes, each consisting of a unitary
    operation applied to $\ket{0}$ state . Each unitary
    is parameterized by a random, fixed vector (\textcolor{grey}{grey}) and the classical
    input. The resulting tensor product state is then measured
    in a fixed basis, and the outcomes are processed
    further by classical post-processing (\textcolor{darkred}{red}) to 
    generate a classical output.
    This procedure can be engineered to approximate any function of the
    classical input.\label{fig:fig2}}
\end{figure}

The quantum kitchen sinks (QKS) algorithm proposed by~\citet{Wilson2018} leverages the connection between trigonometric functions and qubit rotations to implement the required non-linearities. Concretely, the original proposal used \emph{linear} random mixtures of features as angles for single-qubit rotations, optionally followed by entangling operations. A quantum circuit composed of such operations impacts the amplitudes of a quantum state with cosines and sines of weighted mixtures of classical data.

More specifically, let $\mathbf{x} \in \mathbb{R}^p$ be a $p$-dimensional feature vector. Let
$\boldsymbol\Omega = (\omega_1, \ldots, \omega_q)^T$ be a randomized matrix, where for each $k$, $\omega_k$ is a $p$-dimensional vector with $r\leq p$ elements having random values and remaining elements being exactly zero. We can also specify a random $q$-dimensional bias vector $\boldsymbol\beta$. Then we get our set of random quantum circuit parameters $\boldsymbol\theta = \boldsymbol\Omega \, \mathbf{x}+\boldsymbol\beta$.

For a sufficiently large count $E$ interpreted as the \emph{number of episodes} we repeat this randomized synthesis $E$ times to form a set of \emph{encoding parameters} $\{\boldsymbol\Omega_e,\boldsymbol\beta_e\}_{e=1}^E$. This set of parameters is drawn only once and becomes a permanent part of a QKS solution. The $\boldsymbol\Omega_e$ and $\boldsymbol\beta_e$ elements can be drawn from various statistical distributions---e.g., normal, uniform, etc.---and parameters of these distributions become hyperparameters of the method. Choice of variances, for one, has a strong impact on the outcomes---as shown by a cross-validation grid-search in the supplementary material.

Once we have encoded the data for an episode $e \in [E]$ as vectors of circuit parameters
 $\boldsymbol\theta_e(\mathbf{x}) = \boldsymbol\Omega_e \, \mathbf{x} + \boldsymbol\beta_e, \, \mathbf{x} \in {\tt Learning data}$ we need to choose an ansatz for quantum circuit(s) driven by these parameters. As shown by~\citet{Wilson2018} the circuit structure can be quite simple provided that the corresponding unitary transformation depends strongly on the $\boldsymbol\theta$s. For example, the one-qubit ansatz presented in \hyperref[fig:fig2]{Fig. 2} has been proven to work well in multiple datasets. As suggested by the figure, the quantum step starts with some basic quantum state, e.g., $\ket{0}$, and the circuit is followed by a full measurement that extracts a classic bit string and collapses the state. This concludes the quantum feature pre-processing step. The bits extracted by measurement of multiple episodes form an aggregated feature vector that is then fed into a classical linear classification algorithm.

The critical difference between QKS and classical RKS is that in the case of QKS, the aggregated feature vector is by itself stochastic even though the encoding parameters $\{\boldsymbol\Omega_e,\boldsymbol\beta_e\}_{e=1}^E$ are fixed. This is due to the stochastic nature of quantum measurement. Therefore it might be crucial to allow multiple runs (referred to as \emph{shots}) of the same circuit within each episode and average measurement results across these runs. Such a multi-shot approach provides a more accurate representation of non-linearities induced by the quantum encoding. In either case, the totality of quantum steps generates an aggregated feature vector that is post-processed further. \citet{Wilson2018} used Logistic Regression (LR) for the classical post-processing, while here, we explore several different approaches (both classical and quantum) to achieve better performance.

\textbf{Coherently processing quantum kitchen sinks with tensor
  networks}. Although the non-linearities provided by QKS are
sufficient to approximate arbitrary functions of the classical inputs
(and may therefore be used for a wide range of machine learning
tasks), it is not apparent how to optimally engineer multi-qubit
episodes, much less how performance may depend on the number of
qubits in an episode. The results of \citet{Wilson2018} indicated some
potential improvement with the number of qubits per episode, but also
considered only fixed instances of arbitrarily chosen multi-qubit
circuits. In this section, we explore an approach to address this
question by considering coherent processing of the output of the
QKS with highly structured shallow quantum circuits: tree tensor
networks (TTN). We chose to focus on TTNs due to computational
convenience, but expect
additional gains may be possible with other TN structures.

\begin{figure}[!ht]
  \begin{center}
    \includegraphics[scale=0.75]{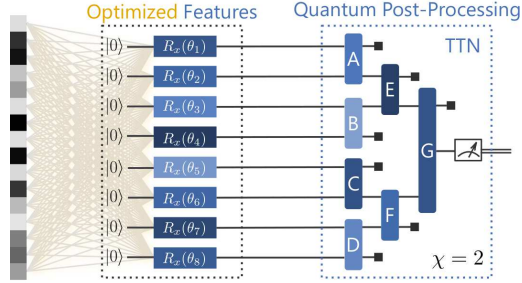}
  \end{center}
  \caption{Coherent post-processing of the quantum features with a 
    tree tensor network (TTN), highlighted in \textcolor{lightblue}{blue}. The 
    TTN is a variational quantum circuit consisting of unitary interactions 
    between qubits. In the $\chi=2$ case, a unitary---represented by a uniquely 
    parameterized tensor---operates on two qubits. Only one of the resulting 
    qubits continues onto the next layer, while the other is measured and 	   
    discarded (as depicted by the black squares). Furthermore, in a departure 
    from the pure randomization approach of QKS, here we also
    consider optimizing the features (\textcolor{gold}{yellow}) as well as the 	
    TTN, using randomization only to initialize the QKS.}
  \label{fig:fig3}
\end{figure}

The original QKS proposal used tensor product measurements, followed
by (largely) unstructured classical post-processing of measurement
outcomes. Here, we wanted to consider measurements that have an
efficiently contractible TN structure
(\hyperref[fig:fig3]{Fig. 3}). This change resulted in a variational
quantum circuit with multiple layers of coherent processing. The
architecture design consisted of two main modules: the non-local
features and TTN.

A TTN will map $E/\lg \chi$ qubits down to $n$ qubit in $O(\lg E)$
depth by successively discarding (or rather, ignoring) half of the
qubits. Note that the TTN we employ here is a slight modification of
the TTN considered in condensed matter
studies~\cite{Shi2006,Tagliacozzo2009,Murg2015,Nakatani2013}. In
particular, our TTN is not made up of isometries (unitaries followed
by projections on some of the outputs) because that would correspond
to post-selection on measurement outcomes at each tensor and an
exponentially small probability of post-selection success for the
overall circuit~\footnote{This may be acceptable in networks that aim
to train {\em generative models}, which would be employed by running
the networks backward and treating the measurements in the isometries
as state preparation. However, such an approach cannot be directly employed
with QKS.}. Instead, in our TTN we
discard some of the output qubits for each tensor in the tree, so
in some sense, our TTN is ``dissipative'', with the advantage that it
does not require post-selection. This changes the connectivity of the
network compared to the isometric case. However, it only translates to an
effective increase in the bond dimension, which is why the contraction
cost has asymptotic scaling proportional to $\chi^7$ instead of the
usual $\chi^3$ for isometric TTN (see details on computational complexity in the supplementary material).

The $O(\chi^7)$ scaling can be understood by noting that the TTN can
be contracted efficiently starting at the leaves. Episodes interact via
$\chi^2\times\chi^2$ unitaries (two input indices and two output
indices, which are then doubled because we consider density matrices),
and a partial trace must be computed to discard half of the episodes at
the output of the unitary. Overall this computation requires summing over four
input indices, two output indices, and a final index for the degrees of
freedom that are traced out, for a total of $7$ indices of dimension
$\chi$. After the leaves are contracted with the first layer of
unitaries, the TTN structure is recovered, so the next layer can be
contracted with the same complexity. This does require a number of
contractions that is linear in $E$, but parallelism allows for $O(\lg
E)$ time contraction.

A quantum computer may implement the contraction by decomposing the
$\chi^2\times\chi^2$ unitary into qubit interactions. This can be done
in depth $O(\chi^4)$ according to the Solovay-Kitaev
theorem~\cite{NielsenChuang00}, pointing to a potential cubic speed-up
over the classical implementation. The more meaningful (and subtle)
comparison to other classical classification algorithms in terms of
the scaling necessary to achieve a fixed target classification
accuracy is not addressed here and remains an open problem.

\textbf{Architecture Design \& Training}. In the proposed architectural
design, each input datum $\mathbf{x}$ is translated into a product state of the form
$
\hat{\rho}_{0}\left(\vec{x}_{n}\right) \otimes \hat{\rho}_{1}\left(\vec{x}_{n}\right) \otimes \hat{\rho}_{2}\left(\vec{x}_{n}\right) \otimes \cdots \otimes \hat{\rho}_{E}\left(\vec{x}_{n}\right)
$
where $E$ is the number of episodes. If we average over all training data
in a class $\ell$, we obtain the separable state
$
\hat{\rho}^{\ell}=\frac{1}{N_{\ell}} \sum_{n=0}^{N_{\ell}} \hat{\rho}_{0}\left(\vec{x}_{n}^{\ell}\right) \otimes \hat{\rho}_{1}\left(\vec{x}_{n}^{\ell}\right) \otimes \hat{\rho}_{2}\left(\vec{x}_{n}^{\ell}\right) \otimes \cdots \otimes \hat{\rho}_{E}\left(\vec{x}_{n}^{\ell}\right)
$
where $N_{\ell}$ is the number of training examples with class label
$\ell$. In the following, we assume binary classification with $\ell
\in \{0,1\}$ for simplicity. However, all the equations can be
straightforwardly generalized for a multi-class setting.

\textit{Tensor network optimization}. We may consider training the TTN
by first fixing the QKS parameters. The unitary tensors are optimized
by gradient descent methods applied to the training
objective described below and in the supplementary material. 
Similar to~\citet{Huggins2019}, the unitaries
are expressed as matrix exponentials of anti-hermitian matrices, while
anti-hermiticity (and thus unitarity) is preserved by an appropriate
choice of parameterization.

We optimize the TTN by minimizing $\Pr(\text{error}) = 1 -
\sum_{\ell\in L} \tr\hat{\rho}^\ell \hat{M}_\ell/|L|$, the probability
of single-shot classification error in the training set, where $L$ is
the set of labels and the $\hat{M}_\ell$ are the elements of a
positive operator-valued measure (POVM). The $\hat{M}_\ell$ are
implicitly defined by the TN and fixed projectors at the
root of the tree TN, i.e., $\hat{M}_\ell =
\mathcal{E}^\dagger(\ketbra{\ell}{\ell})$ for some completely-positive
trace-preserving (CPTP) map $\mathcal{E}$ corresponding to the state
evolution in the dissipative TN. For simplicity, we use
the same objective for single-shot and multi-shot
classification\footnote{The performance in the multi-shot case is not
determined directly by the objective, but rather by numerically
training and testing a linear classifier on the observed outcomes
frequencies for multiple experimental shots on each training/testing
example (i.e., 2 dimensional real-valued vectors)}, although this
approach requires modification for multi-class multi-shot
classification.

Due to the normalization condition on the POVM, binary classification
yields $\Pr(\text{error}) =
\frac{1}{2}-\frac{1}{2}\tr\left[(\hat{\rho}_0-\hat{\rho}_1)\mathcal{E}^\dagger(\ketbra{0}{0})\right]$,
so that we may take the maximization of
$f=\tr\left[(\hat{\rho}_0-\hat{\rho}_1)\mathcal{E}^\dagger(\ketbra{0}{0})\right]$
as our objective instead (with similar expressions for the multi-class
case). It is inconvenient to manipulate $\hat{\rho}_\ell$ directly, as
these matrices have size that is exponential in $E$, but
$c_i(\bar{\ell},\ell)=\tr\left[\hat\rho_i^{\bar{\ell}}\mathcal{E}^\dagger(\ketbra{\ell}{\ell})\right]$
can be computed by contracting the TN for any given
training example $\hat{\rho}_i^{\bar{\ell}}$, which can be done on a
classical computer by only manipulating tensors of a fixed dimension
in $O(\chi)$ for $(\chi\ge|L|)$. On a quantum computer, $c_i(\bar{\ell},\ell)$ may be estimated by running the state preparation and TN as
circuits, and estimating the probability of obtaining outcome $\ell$
at the final measurement for state preparation
$\hat{\rho}_i^{\bar{\ell}}$.

Computing the objective function requires contracting the entire
TN for each training example, which is excessive in some
cases. This can be replaced by considering only random mini-batches of
the training set using stochastic variants of gradient descent (see supplementary material). Suppose we choose to optimize each tensor sequentially. 
In that case, we may partially contract the TN so that at each step of the
optimization, the objective reduces to the product of three small
matrices---the unitary tensor, its conjugate, and a non-unitary tensor
corresponding to the partial contraction of the remainder of the
network which remains fixed---in a manner that is reminiscent of {\em
  quantum combs}~\cite{Chiribella08}. However, we find that
globally updating all tensors leads to faster convergence and lower
computational overhead. 

\textit{Feature optimization}. In a departure from the pure randomization 
approach of kitchen sinks, here we also consider optimizing the features as well as the tensor network, using randomization only to initialize the QKS. 
This additional optimization step can also be approached with gradient descent. For an episode $e \in [E]$, we computed the gradient with respect to the global objective $f$ of the ansatz,
$
\nabla f=\left[\frac{\partial f}{\partial \omega_{1}}, \frac{\partial f}{\partial \omega_{2}}, \cdots \frac{\partial f}{\partial \omega_{d}}, \frac{\partial f}{\partial \beta}\right],
$
with respect to parameters $\{\mathbf{\Omega}_{e}, \boldsymbol\beta_{e}\}$ of the
driving angles $\mathbf{\theta}_{e}(\mathbf{x})=\mathbf{\Omega}_{e} \mathbf{x}+\boldsymbol\beta_{e}$. For a batch of examples indexed by $n \in [N_{\ell}]$, the gradient elements are
$
\frac{\partial f}{\partial \omega_{i}}=\frac{1}{N_{\ell}} \sum_{n}\left(\frac{\partial \rho_{n}}{\partial \theta_{n}} x_{n, i}\right) \cdot V_{n}$ and $\frac{\partial f}{\partial \beta}=\frac{1}{N_{\ell}} \sum_{n} \frac{\partial \rho_{n}}{\partial \theta_{n}} \cdot V_{n}$ where $\rho_n$ represents the density matrix of the feature being optimized and ``$\cdot~V_n$'' represents contraction with the remainder of the tensor network. With the gradient in hand, we sequentially optimize each feature with respect to the global objective. Additional architecture and training details can be found in the supplementary material.

\textbf{Reducing model complexity via translational symmetry and sparsity constraints}. We explore two approaches for reducing model complexity by imposing (\textit{i}) translational symmetry on the TTN (see motivation in the supplementary material) and (\textit{ii}) sparsity on the features. Translational symmetry is enforced by requiring all tensors at a fixed distance from the root to be identical, while sparsity is enforced by setting a fixed (random) fraction of $\boldsymbol\Omega_{e}$ to 0. These constraints reduce the size of the hypothesis space and could offer better generalization, that is, up to a point where the model begins to underfit. Parsimonious models with minimal complexity could exist right before this transition.

\begin{figure}[h!]
  \begin{center}
    \includegraphics[scale=0.75]{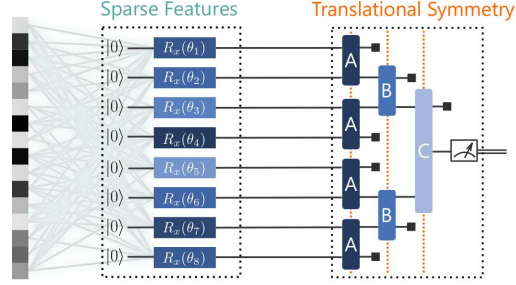}
  \end{center}
  \caption{Translational symmetry and sparsity may be
    imposed on the TN and features, respectively, thereby reducing the
    number of parameters in the model. Note the sparser connectivity
    between data features and quantum rotations (\textcolor{teal}{teal}), and the
    layer-wise symmetry of the tensors (\textcolor{orange}{orange}).}
  \label{fig:fig4}
\end{figure}

\section{Results}
\label{sec:results}
In this section, we present the results of several numerical experiments on QKS, 
QKS+TTN, and QKS+TTN+FO. We begin 
by establishing the baseline performance of QKS. Next, we combine QKS with a TTN---we found that simply combining QKS and TTN did not improve classification 
performance over the QKS baseline. However, the addition of 
feature optimization greatly improved the performance and yielded state-of-the-art quantum circuits for image classification. Finally, 
for the most complex networks, we show that imposing translational symmetry and 
sparsity constraints led to dramatic reductions in the number of free parameters 
while having minimal impact on classification performance.

\textbf{Baseline performance of quantum kitchen sinks}. The original QKS design, as described by~\citet{Wilson2018}, was intentionally limited to circuits where the quantum encoding of the data features was fed into a linear post-processing layer. For the (3,5)-MNIST dataset, error rates between 3.3\% and 3.7\% were reported in the experiments that ran on noisy quantum hardware. Results obtained on a noiseless simulator showed error rates at or slightly below 2\%. While these error rates showed a lift over a linear SVMs (which was the stated intent of the original QKS research), the performance has been inferior compared to the state of the art for (3,5)-MNIST classification.

\begin{figure}[h!]
  \begin{center}
    \includegraphics[width=\textwidth]{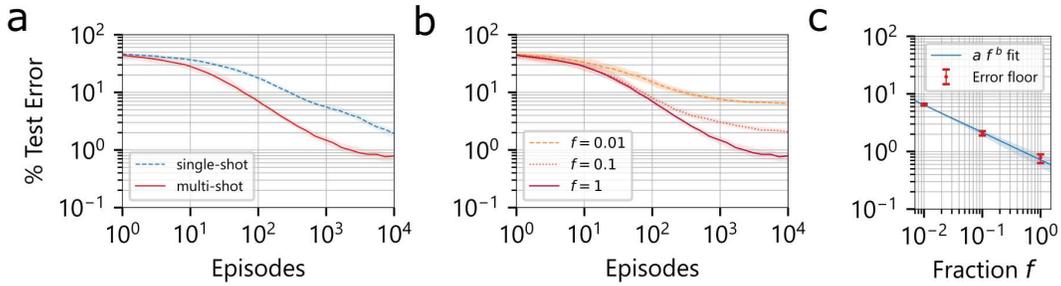}
  \end{center}
  \caption{(a) Classification error rates for (\textit{3,5})-MNIST over 100 
    realizations (median and
    68\% credibility intervals) as a function of the number of QKS episodes. 
    Allowing multiple shots at the one-qubit ansatz (multi-shot) led to 
    better performance compared to single shots at the same circuits 
    (single-shot). (b) Restricting the training dataset to a fraction of its 
    original size ($f=0.1$ and $f=0.01$) limited the achievable classification 
    performance, and (c) the observed dependency on the
    size of the training dataset was fit to a power law $y(f)=a~f^b$ where
    $(a,b) = (0.72 \pm 0.18, -0.48 \pm 0.06)$, consistent with the
    $O(1/\sqrt{N})$ predictions from theoretical
    results~\cite{Rahimi07,Rahimi08a,Rahimi08b}.}
  \label{fig:fig5}
\end{figure}

We established a more competitive baseline than the one given in the original QKS proposal by allowing multiple runs, or \textit{shots}, of the quantum circuits. Compared to the single-shot approach, the multi-shot approach offered better error rate scaling as a function of the number of episodes (\hyperref[fig:fig5]{Fig. 5a}). At the largest number of episodes we tested ($E=10,000$), the single-shot approach yielded a test error of $1.87 \pm 0.28$ (mean $\pm$ std.) on \textit{(3,5)-MNIST}, which was consistent previous observations~\cite{Wilson2018}, while the multi-shot approach offered a significant reduction to $0.78 \pm 0.12$.

Further, using the multi-shot approach, we validated the scaling of the error rate as a function of the training set size ($N$) against theoretical guarantees for random feature algorithms. We simulated smaller datasets by taking fractions of the training dataset ($f=0.1$ and $f=0.01$), which shifted the noise floor higher at a large number of episodes \hyperref[fig:fig5]{Fig. 5b}. The empirical dependence of the noise floor on the size of the training set was fitted to a power law $y(f)=a~f^b$ via a non-linear least-squares procedure yielding $(a,b) = (0.72 \pm 0.18, -0.48 \pm 0.06)$,\footnote{The uncertainties are computed by linearization of the fit model in the neighborhood of the least-squares optimum, and correspond to a single standard deviation along each coordinate.} which is consistent with the $O(1/\sqrt{N})$ predictions from theoretical results~\cite{Rahimi07,Rahimi08a,Rahimi08b}. For the full dataset ($f=1$) and a large number of episodes ($E \rightarrow \infty$), the power law predicts an error of $a = 0.72 \pm 0.18$, which is consistent with the error $0.78 \pm 0.12$ observed at $E=10,000$. Note that the one-qubit QKS ansatz corresponds to the random Fourier features of RKS; therefore, we treat QKS+SVM as the classical model and explore improvements against this baseline via coherent post-processing.

\textbf{Coherent processing of non-local features yields compact networks}. From the results presented in the preceding section, it was clear that many episodes were required for QKS to perform well. Up to this point, the quantum states generated by QKS have been measured and classically post-processed with a linear SVM, and the parameters of QKS have been randomly drawn/fixed. Here we have replaced classical post-processing with quantum post-processing---QKS+SVM to QKS+TTN---and optimized the features along with the tensor network (\textit{feature optimization}), using randomization only to initialize QKS. We found that QKS+TTN yielded no qualitative improvements, however, the addition of feature optimization (QKS+TTN+FO) greatly boosted performance.

\begin{figure}[h!]
  \begin{center}
    \includegraphics[width=\textwidth]{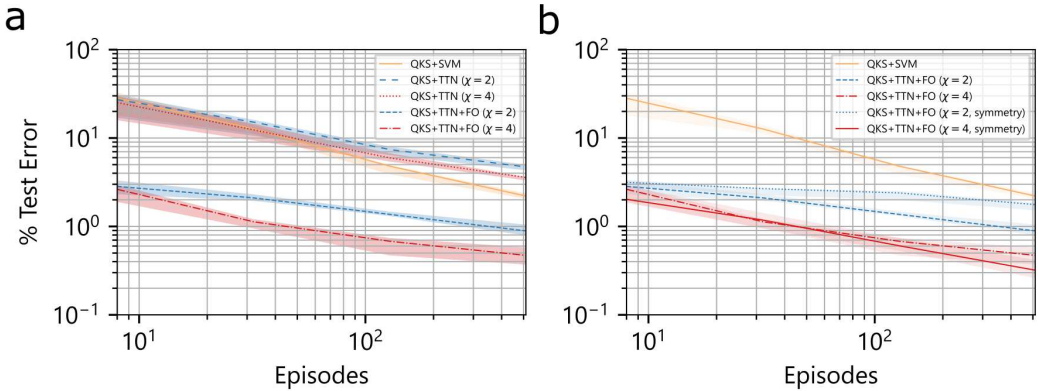}
  \end{center}
  \caption{(a) Performance of combined protocols on (\textit{3,5})-MNIST over 10 realizations (median and 68\% credibility intervals) as a function of the number of episodes or qubits. It shows that feature-optimized networks offer a significant performance lift over the QKS baseline while admitting a $20 \times$
  	reduction in qubit utilization. (b) Imposing translational symmetry constraints on the TN degrades the performance of $\chi=2$ networks but enhances the performance of $\chi=4$ networks.}
  \label{fig:fig6}
\end{figure}

We compared the performance of networks {\em without} feature optimization (QKS+TN) against those {\em with} feature optimization (QKS+TTN+FO).  Networks without feature optimization (QKS+TTN) closely tracked the QKS+SVM baseline while networks with feature optimization (QKS+TTN+FO) outperformed QKS+SVM, while admitting a $20 \times$ reduction in the number of episodes/qubits (\hyperref[fig:fig6]{Fig. 6a}). Therefore, it appears that \textit{feature optimization is necessary for the formation of compact networks}.

\textbf{Reducing model complexity: translational symmetry and sparsity constraints}.  In the previous section, we showed that QKS+TTN+FO achieved better performance than QKS+SVM at a fraction of the number of episodes. We also reduced the model complexity, as measured by the number of free (trainable) parameters by imposing translational symmetry on the TTN and sparsity on the features. 

Translational symmetry constraints on the TTN had varying effects, depending on the bond dimension of the network (\hyperref[fig:fig6]{Fig. 6b}). While performance was degraded in the $\chi=2$ networks, the performance of $\chi=4$ networks was equivalent or slightly better (\hyperref[table:model-reduction]{Table 1})---while admitting an exponential reduction in the number of parameters in the TTN.

\begin{table}[h!]
\caption{Translational symmetry and sparsity constraints were imposed on most complex model we tested, a QKS+TTN+FO network with 512 qubits and $\chi=4$. Imposing these constraints yielded dramatic reductions in the number of free parameters. A combination of translational symmetry and 50\% sparsity constraints yielded the best test errors (mean $\pm$ std.) on (\textit{3,5})-MNIST, in \textbf{bold}.} 
\centering
\begin{tabular}{ccc}
\\
 \hline
 \rule{0pt}{2ex}Model Reduction & Free Parameters (Fixed) & Test Error (\%)\\
 \hline
 \rule{0pt}{2ex}No symmetry or sparsity   &  $466,945$ &  $0.44 \pm 0.13$\\
 Transl. Symmetry     &  $403,960$ &  $0.35 \pm 0.07$\\
 Transl. Symmetry + Sparsity (50\%)  &  $203,256$ &  $\mathbf{0.35 \pm 0.04}$\\
 Transl. Symmetry + Sparsity (25\%)  &  $102,904$ &  $0.37 \pm 0.07$\\
 Transl. Symmetry + Sparsity (13\%)    &  $52,728$ &  $0.45 \pm 0.08$\\
 Transl. Symmetry + Sparsity (6\%)  & $27,640$ &  $0.65 \pm 0.18$\\

\hline
\end{tabular}
\label{table:model-reduction}
\end{table}

Further, in the most complex model we tested, a QKS+TTN+FO network with with 512 episodes and $\chi=4$, imposing both translational symmetry on the TTN and \textit{sparsity} in the features revealed a regime of parsimonious models (\hyperref[table:model-reduction]{Table 1}). The design of the best performing model on (\textit{3,5})-MNIST had a combination of constraints: a TTN with translational symmetry and features with 50\% sparsity. Note, however, that all parsimonious models had test errors below the QKS baseline.

\textbf{State-of-the-art quantum circuits for image classification}. We performed benchmarks on a broader set of classification tasks. On the hardest MNIST binary classification tasks, the non-local feature maps of QKS led to significantly lower test errors compared to local feature maps, when the pre-processing and architecture/training configurations were matched (Table~\ref{table:state-of-the-art}). The best performing QKS+TTN+FO model (with translational symmetry) from the previous section led to the lowest test errors we observed. These performance gains motivated a larger investigation into the full MNIST dataset and the general multi-class problem.

\begin{table}[h!]
\caption{Non-local feature maps led to state-of-the-art quantum circuits for image classification, as demonstrated on the hardest MNIST binary classification problems. We ran a \textit{control} experiment that matched the pre-processing and configuration (64 qubits, $\chi=2$, a mini-batch size of 222, and 30 epochs) of a similar quantum tensor network approach employing local feature maps~\cite{Huggins2019}. Additionally, we tested the \textit{best} performing QKS+TTN+FO model (with translational symmetry), which involved no pre-processing and the following configuration: 512 qubits, $\chi=4$, a mini-batch size of 32, and 40 epochs.}
\centering
\begin{tabular}{P{0.4\linewidth}P{0.07\linewidth}P{0.07\linewidth}P{0.07\linewidth}P{0.07\linewidth}P{0.07\linewidth}}
\\
 \hline
 \rule{0pt}{2ex}Quantum Encoding & 3 vs. 5 & 4 vs. 9 & 7 vs. 9 & 3 vs. 9 & 2 vs. 7\\
 \hline
 \rule{0pt}{2ex}Local feature maps~\cite{Huggins2019}  &  $12.4\%$ &  $12.0\%$ &  $10.7\%$ &  $5.9\%$ &  $4.3\%$  \\
 Non-local feature maps (control) & $5.1\%$ &  $4.2\%$ &  $4.5\%$  &  $2.5\%$ &  $2.2\%$ \\
 Non-local feature maps (best)  &  $0.4\%$ &   $1.1\%$  &  $0.8\%$ &  $0.8\%$ & $0.9\%$\\
\hline
\end{tabular}
\label{table:state-of-the-art}
\end{table}

For the multi-class classification of MNIST, we established a baseline with QKS+SVM by constructing a multi-class model composed of one-versus-one (OvO) binary classifiers: at 10,000 episodes, the multi-class test error was 2.31\%. In comparison, the QKS+TTN+FO with translational symmetry, trained as OvO binary classifiers, yielded a multi-class test error of 1.16\% at 512 episodes/qubits. Taken together with the complete set of classification error rates (see supplemental material), the QKS+TTN+FO offers state-of-the-art quantum circuits for image classification.

\section{Discussion}
\label{sec:discussion}

We have demonstrated that combining the non-local feature maps (of QKS) with coherent processing (using a TN) and feature optimization can yield significant
improvements in classification error rates over the original QKS
proposal, leading to state-of-the-art quantum circuits for image classification
with a qubit count and circuit depth that are within reach of near-term devices.

While these gains do not translate to meaningful quantum advantage (the TTN can
be efficiently contracted in a classical computer), they illustrate
that variational optimization of quantum circuits can perform
non-trivial tasks on relatively small devices. The tree structure of
the TN considered here has also been argued by~\citet{Kim2017} to have
favorable noise-resilience properties, which makes them promising
candidates for interesting near-term demonstrations.

We did not take into consideration the impact of connectivity
constraints of any experimental realization of a quantum computer. It
should be noted, however, that connectivity constrains impose at most
a linear depth overhead as a function of
width~\cite{Bhattacharjee2017}, but more detailed (and likely
architecture-dependent) analysis would be needed to pin down concrete overhead numbers.

Although the classification error rates reported here are state-of-the-art for quantum circuits~\cite{Grant2018,Huggins2019,Farhi2018,Chen2021,Hur2022} and competitive against quantum-inspired tensor networks~\cite{Stoudenmire2016,Stoudenmire2018,Liu2019}, they are comparable to classical methods~\cite{LeCun1998}. Thus, modules of the ansatz would likely benefit from further exploration of different feature maps~\citep{Simard2003} and other tensor network structures~\cite{Orus2019}. The upshot would be a class of efficient unitary networks that could fundamamentally resolve the exploding/vanishing gradient problem in deep neural networks and the detection of long-term dependencies in dynamical systems~\cite{Arjovsky2016,Jing2017}.

\begin{ack}
  NXK acknowledges support from a Microsoft Quantum Internship, where
  this collaboration started.  AB and MPS thank Martin
  Roetteler, Dave Wecker, and Stephen Jordan for fruitful discussions
  and encouragement.
\end{ack}

\bibliographystyle{unsrtnat}
\bibliography{bib-qks-tn}{}


\appendix
\renewcommand\thefigure{\thesection.\arabic{figure}}
\setcounter{figure}{0}
\section{Computational Complexity\label{app:complexity}}

As we have stated in the introduction,
tensor network (TN) methods have been widely studied in the machine learning and
physics communities due to their expressive power and computational
convenience. Under some mild constraints on
the components, a TN may be interpreted as quantum
circuits, and in some cases, these circuits
may be simulated with a computational effort that scales polynomially
with the number of qubits involved---in stark contrast with the
exponential scaling that may be expected for arbitrary
circuits.

A concrete example of a TN that can be simulated
efficiently is a {\em tree tensor network} (TTN). In these networks, two
collections of $n$ qubits each (where $\chi=2^n$ is referred to as the
{\em bond dimension}) interact unitarily~\footnote{We also considered
more general operations, such as quantum channels. Although quantum
channels have an appealing structure for numerical optimization (the
optimization of each local tensor is a convex problem), it did not
appear to lead to any performance improvement over the strictly
unitary case.}, but only one of the resulting collections of $n$ qubits
continues on for additional computation (the other is simply
ignored). By successively discarding half of the qubits, a TTN 
will map $E/\lg \chi$ qubits down to $n$ qubits in $O(\lg E)$
depth. In this work, we only considered $\chi=2$ and $\chi=4$, and
focused on the TTN to build a large coherent
computation.

Classification using QKS+TTN requires simulating the evolution of the
quantum state through the fixed circuits (parameterized by the
classical input data) and the TTN, which concludes
with the application of a measurement on the remaining $n$ qubits at
the root of the tree. The result outcome (or its expectation value)
is then used for classification.

Thanks to the tree structure of the TN, the output of
QKS+TTN can be computed by a classical computer in time
$O(\chi^{7}~E)$ if the TN is contracted serially or in
time $O(\chi^{7}~\lg E)$ if the network is contracted in parallel. A
quantum computer may implement the same transformation in time
$O(\chi^3)$, with the disadvantage that the final outcome can only be
sampled (but this sampling can be done in parallel by running multiple
copies of the circuit). The classical data must be pre-processed for
each episode, with a runtime of $O(E~D)$ serially or $O(D)$ in
parallel, where $D$ is the dimensionality of the classical data.

Training the network, however, requires more than just being able to
simulate the evolution of the quantum state: it requires optimizing
over all the parameters in the TN. Without imposing any
symmetry, the TTN has $(E-1)\chi^2(\chi^2+1)/2\lg\chi$
real parameters. If the tensors along each layer of the TTN
are constrained to be identical, the number of parameters is
reduced to $\lg(E)\chi^2(\chi^2+1)/2\lg\chi$. Overall, the cost of
training will be proportional to the number of parameters in the model
and to the training batch size.

The classical time overhead of classification with QKS+TTN is
$O(D+\chi^{7}~E)$, yielding the full distributions of outcomes.  The
quantum time overhead with the equivalent circuits is $O(S~D+S~\chi^4~\lg E)$,
where $S$ is the number of runs of the quantum circuit (given we
can only sample the final measurement in the quantum case, instead of
obtaining the full distribution). The time overhead for the training of
QKS+TTN picks up additional factors proportional to the number of
parameters in the TN, but these are the same for the
classical and quantum cases.

While for $\chi>2$, the compilation of each tensor into the native
gate operations may add significant classical computation overhead
during training and significantly increase circuit depth, one may
replace the monolithic $\chi>2$ tensors with a network of
subcomponents of bond dimension $2$. Such replacement would require a
number of tensors/unitaries that grows as $O(\chi^4)$ in order to
achieve good accuracy in full generality~\cite{NielsenChuang00}, but
one may consider instead the replacement with a fixed network of
subcomponents, and optimizing over the parameters of that fixed
network~\cite{Haghshenas2022}. Although we do not consider the impact
of such a fixed structure here, there are indications that it would
not impact classification performance---with the advantage that it
would greatly reduce the number of free parameters in the network, and
the TN optimization would be cheaper than computing accurate unitary
decompositions at every iteration.

As pointed out in the main text, the TTNs we employ
here are a slight modification of the TTNs considered
in condensed matter studies. In particular, our TTN is
not made up of isometries.  Instead, in our TN we
``ignore'' or ``discard'' some of the output qubits for each tensor in
the tree, so in some sense, our TTN is ``dissipative.''
This changes the connectivity of the network compared to the isometric
case, but it only translates to an effective increase in the bond
dimension, which is why the contraction cost has asymptotic scaling
proportional to $\chi^7$ instead of the usual $\chi^3$ for isometric
TTNs.

\newpage
\section{Materials and Methods\label{app:materials-methods}}

\subsection{Architecture and Training Details\label{app:training-details}}
\textit{Quantum kitchen sinks}. We simulated quantum kitchen sinks (QKS) using NumPy~\cite{Harris2020} and processed the classical outputs using scikit-learn~\cite{scikit-learn}. Hyperparameters such as the variance of QKS weights and the regularization parameter of linear SVMs were optimized using cross-validation grid search and randomized search, respectively. For each experimental condition, we collected 32 logarithmically spaced sample points and 100 realizations using Dask~\cite{Rocklin2015} to distribute the computation across an internal cluster with 340 nodes, where each node had 44 cores and 318 GB.

\textit{Tree tensor networks}. We combined QKS with a tree tensor network (TTN) by using TensorNetwork~\cite{Roberts2019} to construct/contract the networks and SciPy~\cite{2020SciPy} to initialize/optimize the tensors and features. The TTN was initialized by drawing $N$ tensors from a circular unitary ensemble (CUE)~\cite{Dyson1962}---$U_n \sim \mathsf{CUE} (\chi^2)$ for $n \in [N]$, where the eigenvalues of $U_n$ have unit length and uniformly distributed in phase---then, for each unitary tensor, we computed the $\chi^2-1$ real numbers parameterizing the corresponding hermitian matrix $H_n$ in $U_n = \exp(iH_n)$. To train the TTN, we chose to optimize each tensor sequentially---below, we discuss global updates to all tensors. In that case, we partially contracted the TTN so that at each step of the optimization, the objective was reduced to the product of three small matrices---the unitary tensor, its conjugate, and a non-unitary tensor corresponding to the partial contraction of the remainder of the network which remained fixed---in a manner that was reminiscent of \textit{quantum combs}~\cite{Chiribella08}. The partial contractions were computed and summed over a batch of $B=1024$ examples---an embarassingly parallel computation that was distributed using Dask. This distributed operation resulted in a small matrix that could be used as a fixed input to any number of optimization methods. To optimize the corresponding unitary tensor, we used conjugate gradient (with a tolerance of $10^{-5}$ and a maximum number of iterations set at 100) since it led to the fastest convergence.

\textit{Feature optimization}. The sequential approach to optimizing each tensor was extended to tuning each feature, which we called feature optimization (FO). The intuition behind FO was derived from classical work by Bengio and LeCun comparing kernel machines and neural networks, where the key difference was whether the low-level features (or \textit{templates matching units}) were tunable~\cite{BengioLeCun2007}. It was argued that more compact networks could be formed by tuning the basis functions in the features to the training objective. Since the combination of QKS+TTN did not offer a performance boost over the QKS+SVM baseline, we hypothesized that compact networks could be formed by optimizing the features---i.e., QKS+TTN+FO could lead to performance gains over the QKS+SVM baseline. Note, the propagation function (or input currents) to a unit $i$ in the the first hidden layer of a neural network $z_i = W_i x + b_i$  is equivalent to the driving angle $\mathbf{\theta}_{e}(\mathbf{x})=\mathbf{\Omega}_{e} \mathbf{x}+\boldsymbol\beta_{e}$ of a QKS episode. However, while the input currents are acted upon by sigmoid/rectifier activation functions in the neural network, the driving angles parameterize a qubit rotation in the case of the one-qubit QKS ansatz.

Using randomization only to initialize features, the addition of feature optimization (QKS+TTN+FO) involved several changes to the optimization of the unitary tensors. For an episode $e \in [E]$, we computed the gradient with respect to the global objective $f$ of the ansatz,

$$\nabla f=\left[\frac{\partial f}{\partial \omega_{1}}, \frac{\partial f}{\partial \omega_{2}}, \cdots \frac{\partial f}{\partial \omega_{d}}, \frac{\partial f}{\partial \beta}\right],$$ 

with respect to parameters $\{\mathbf{\Omega}_{e}, \boldsymbol\beta_{e}\}$ of the
driving angles $\mathbf{\theta}_{e}(\mathbf{x})=\mathbf{\Omega}_{e} \mathbf{x}+\boldsymbol\beta_{e}$. For a batch of examples indexed by $n \in [N_{\ell}]$, the gradient elements are

$$
\frac{\partial f}{\partial \omega_{i}}=\frac{1}{N_{\ell}} \sum_{n}\left(\frac{\partial \rho_{n}}{\partial \theta_{n}} x_{n, i}\right) \cdot V_{n}, \quad \frac{\partial f}{\partial \beta}=\frac{1}{N_{\ell}} \sum_{n} \frac{\partial \rho_{n}}{\partial \theta_{n}} \cdot V_{n},$$

where $\rho_n$ represents the density matrix of the featuring that is being optimized and ``$\cdot~V_n$'' represents contraction with the remainder of the tensor network. Similar to the optimization of the TTN, conjugate gradient was used for feature optimization, where the only change was that the maximum number of iterations was set at 5. For different bond dimensions ($\chi=2$ and $\chi=4$), we compared QKS+TTN and QKS+TTN+FO at 500 sweeps by collecting 4 linearly spaced sample points and 10 realizations using Dask on the internal cluster (described above).

\textit{Translational symmetry and sparse features}. Reducing the complexity of a learning model is directly related to its generalizability and speed of convergence~\cite{LeCun1989}. While there is a well-known 2D translational symmetry that may be exploited for image classification~\cite{Simard2003, LeCun1989}, it was not considered in this work; rather, we focused on the permutation symmetry in the independent and identically distributed QKS episodes, with the goal of reducing the number of trainable parameters in the TTN. 

Since translation symmetry is included in the permutation symmetry, we imposed layer-wise weight sharing on the TTN. Note, FO breaks the symmetry of the QKS episodes---however, reducing number of parameters by imposing translational symmetry is still well-motivated, because such a constraint is beneficial in avoiding barren plateaus~\cite{Wang2020, Cerezo2021, Arrasmith2021} and overfitting~\cite{LeCun1989}. In short, we did not exploit the 2D translational symmetry in the images, but instead we were motivated by the permutation symmetry in the distribution of QKS episodes before FO. Imposing translational symmetry required a different training algorithm since each layer in the TTN shared the same set of parameters. We implemented a version of the architecture in TensorFlow~\cite{tensorflow2015} so that we could make global updates via automatic differentiation (Algorithm~\ref{alg:alg1}). All parameters were updated using Adam~\cite{Kingma2015}, similar to Grant et al.\cite{Grant2018}, a batch size of $B=32$, and an initial learning rate of $0.001$. We compared QKS+TTN+FO with and without translational symmetry in the most complex model we tested ($E=512$ and $\chi=4$) at 40 epochs and over 10 realizations using a single desktop with 4 cores and 32 GB.

Using the Tensorflow implementation, we also imposed sparsity in the features by setting a fixed (random) fraction of $\boldsymbol \Omega_e$ to zero. We sought the most parsimonious QKS+TTN+FO models (for $E=512$ and $\chi=4$) by sampling 5 exponentially decreasing densities $d$ at 40 epochs and over 10 realizations using the single desktop  (described above).

\begin{algorithm}
\caption{Training algorithm for QKS+TTN+FO}\label{alg:alg1}
\begin{algorithmic}[1]
\State{\textbf{given} training data $\{\mathbf{x}_m,\ell_m\}_{m=0}^M$, batch size $B$, number of QKS episodes $E$, sparsity of features $d \in (0,1]$, bond dimension $\chi$, number of tensors $N$, and translational symmetry flag $f_{sym}$.}\\

\State{\textbf{initialize} random parameters $\boldsymbol \Omega_e \sim \mathcal{N}(0,\sigma^2)$ and $\boldsymbol\beta_e \sim \mathcal{U}(0,2\pi)$ for $e \in [E]$, and Haar random unitaries $\{U_n\}_{n=1}^N$, where $U_n \sim$ CUE$(\chi^2)$.}\\
\\
$\Theta \leftarrow (\{\boldsymbol\Omega_e,\boldsymbol\beta_e\}_{e=1}^E, \{U_n\}_{n=1}^N)$\\

\Function{VariationalCircuit}{$\Theta$}
	\ForAll {$e \leq E$}
		\State{$\boldsymbol \theta_e \leftarrow \mathbf{\Omega}_{e} \mathbf{x}_m+\boldsymbol\beta_{e}$}
		\State{$\rho_e$ $\leftarrow$ \textsc{PrepareFeature}($\boldsymbol \theta_e$, $d$)}
	\EndFor
	\State{$f \leftarrow$ \textsc{ContractTTN}($\{\rho_e\}_{e=0}^E$, $\ell_m$, $\chi$, $\{U_n\}_{n=1}^N$, $f_{sym}$)}
	\State{\Return $f$}
\EndFunction\\

\Repeat
	\State{$\nabla f(\Theta) \leftarrow$ \textsc{GradientOnBatch}(\textsc{VariationalCircuit}($\Theta$), $B$)}	
	\State{$\Theta \leftarrow$ \textsc{UpdateParameters}($\nabla f(\Theta)$)}
\Until{\textit{stopping criterion is met}}\\
\Return trained classifier \textsc{VariationalCircuit}($\Theta$)
\end{algorithmic}
\end{algorithm}

\textit{Multi-class classification}. We applied the previously described architectures to a broader set of binary and multi-class classification problems by training many one-versus-one (OvO) classifiers. We tested a more general alternative that adapted the QKS+TTN+FO ($\chi=4$) architecture for the multi-class problem by mapping the full 16-dimensional output at the root node to 10 classes. We found that mapping 12 of the measurement outcomes to 6 classes and the remaining outcomes to 4 classes led to the fastest convergence; however, more general mappings may be considered.

All benchmarks for binary classification did not involve any pre-processing or regularization techniques. For the multi-class classification of MNIST, we added a pre-processing step that deskewed MNIST images. Additionally, to reduce overfitting and improve convergence, we replaced Adam with AdamW with warm restarts, where the learning rate and decoupled weight decay (initialized at $0.001$ and $4 \times 10^{-4}$) followed cosine annealing schedules (5 restarts, $T_0=1$, $T_{mult}=2$)~\cite{Loshchilov2018}. 
\subsection{Licensing of Existing Assets\label{app:licensing}}
We provide licensing information for each existing asset below:

\textbf{Python Libraries}
\begin{itemize}
  \setlength\itemsep{0em}
  \item NumPy \cite{Harris2020} and SciPy~\cite{2020SciPy} are BSD licensed.
  \item Dask~\cite{Rocklin2015} and scikit-learn~\cite{scikit-learn} are licensed under the New BSD License.
  \item TensorFlow~\cite{tensorflow2015} and TensorNetwork~\cite{Roberts2019} are licensed under the Apache License 2.0.
\end{itemize}

\textbf{Datasets}
\begin{itemize}
  \setlength\itemsep{0em}
  \item MNIST~\cite{LeCun1998} is licensed under the Creative Commons Attribution-Share Alike 3.0 
  license.
  \item Fashion-MNIST~\cite{Xiao2017} is licensed under the MIT license.
\end{itemize}

\newpage
\section{Supplementary Results\label{app:supplementary-results}}

\subsection{Hyperparameters for Quantum Kitchen Sinks\label{app:qks-hyperparameters}}
\begin{figure}[h!]
  \begin{center}
    \includegraphics[width=\textwidth]{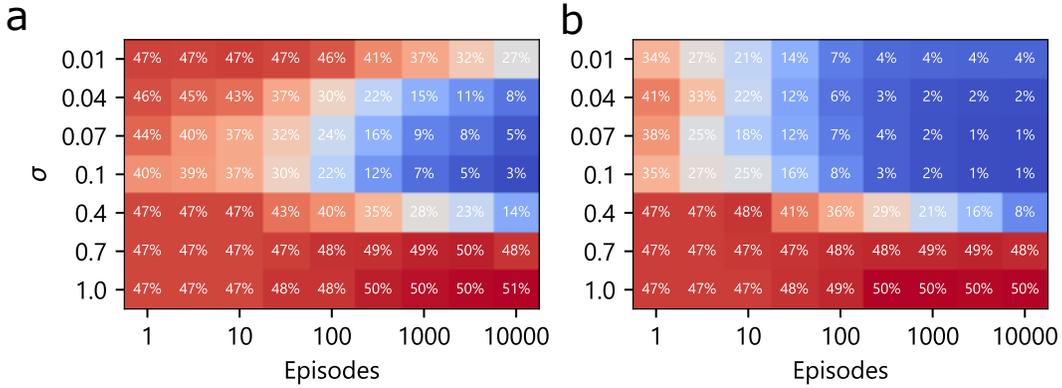}
  \end{center}
  \caption{(a) Single-shot validation errors (averaged over 5 folds) indicate optimal hyperparameter values: $\sigma=0.1$ and $E=10,000$. (b) The same hyperparameters hold for the multi-shot case.}
  \label{fig:figS1}
\end{figure}

Several hyperparameters in quantum kitchen sinks (QKS) affected its classification performance: the variance of the normal distribution from which the weights of the random features are drawn, controlled by $\sigma$, and the number of episodes $E$. We performed cross-validation (5 folds) experiments using a grid search over these hyperparameters for both single-shot and multi-shot circuits (Figure~\ref{fig:figS1}) and found an optimal value of $\sigma=0.1$ led to the best validation errors at a large number of episodes ($E=10,000$).

\newpage
\subsection{Ablation of the Tensor Network\label{app:feature-enrichment}}
\begin{figure}[h!]
  \begin{center}
    \includegraphics[width=\textwidth]{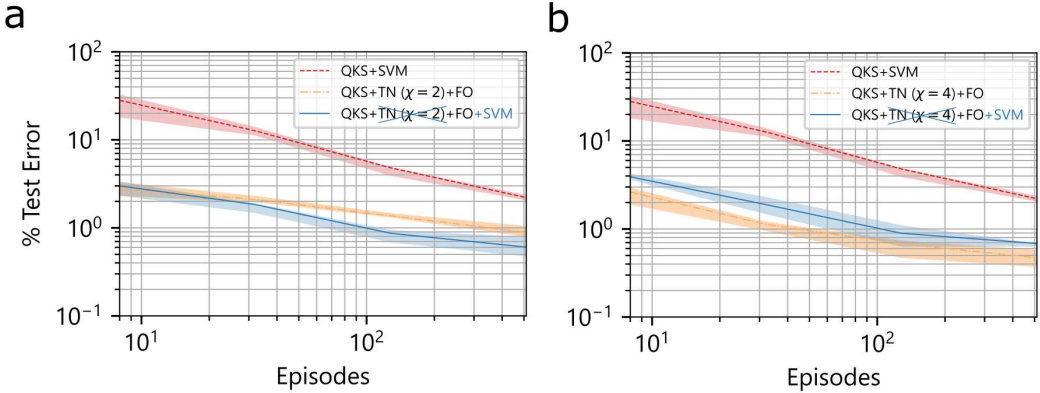}
  \end{center}
  \caption{(a) In the $\chi=2$ network, the relative contribution of
    feature enrichment exceeded that of coherent post-processing and
    the reduction of the feature representation by tensor network led
    to a degradation in performance. (b) For the $\chi=4$ network,
    there is a near-equivalent contribution of feature enrichment,
    however, the coherent post-processing of the tensor network offered
    a performance boost over classical post-processing.}
  \label{fig:figS2}
\end{figure}

In the main text, we showed that the addition of feature optimization (FO) to QKS+TTN was necessary for significant improvements over the QKS+SVM baseline. An important question is whether the improvements were solely due to the enrichment of the non-local features, which we refer to as \textit{feature enrichment}. Moreover, how much of the improvement could be attributed to the \textit{coherent post-processing} with tensor networks? We assessed the  contribution of these two factors by removing the tensor network after training, and replacing the tensor network with classical post-processing (a linear SVM classifier) on the optimized non-local features (Figure~\ref{fig:figS2}).

We observed different behavior for different bond dimensions. For $\chi=2$,
classical post-processing the enriched features outperformed coherent
post-processing of the enriched features, despite the features having been
optimized for the coherent post-processing of the TTN. However, for
$\chi=4$, classical post-processing of the enriched features
under-performed coherent post-processing of the enriched features. Interestingly, the
performance of classical post-processing was essentially the same in
these two cases, while the coherent processing improved by a factor of
two with the larger bond dimension.  This provides some indication that
both feature enrichment and coherent post-processing contribute to the
performance improvement of QKS+TTN+FO over QKS+TTN and QKS+SVM.  In future work, this
question can be more thoroughly investigated by studying feature
optimization with purely classical post-processing and by exploring
different tensor network features in coherent post-processing
(different connectivity, higher bond dimensions, etc.).

\newpage
\subsection{(Top, Shirt)-Fashion-MNIST\label{app:fashion-mnist}}

We also considered the performance of QKS+SVM, QKS+TTN, and QKS+TTN+FO (with and without translational symmetry) on the Fashion-MNIST~\cite{Xiao2017} dataset. The Fashion-MNIST dataset~\cite{Xiao2017} is composed of $28 \times 28$-pixel, 8-bit grayscale images, each depicting various merchandise and articles of clothing, making it a more difficult classification task. We vectorize and split the dataset similarly to that of MNIST. We tested the models on a hard binary classification problem, ``Top'' vs. ``Shirt,'' which we refer to as \textit{(Top vs. Shirt)-Fashion-MNIST}.

\begin{figure}[h!]
  \begin{center}
    \includegraphics[width=\textwidth]{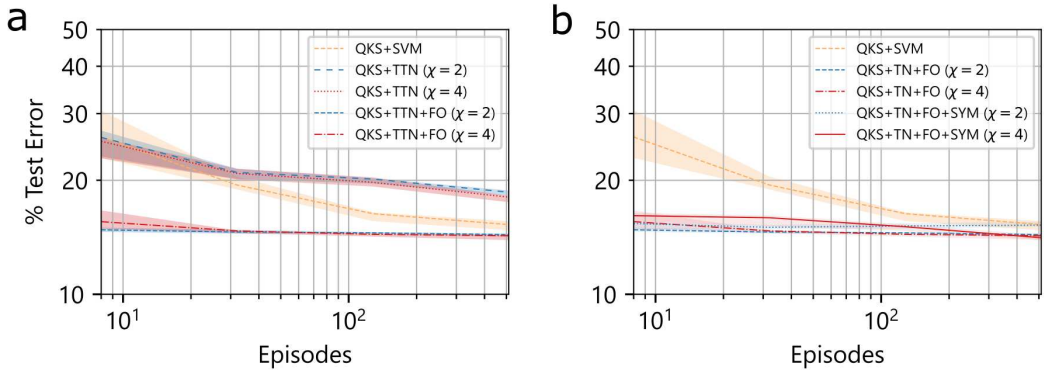}
  \end{center}
  \caption{(a) Classification error rates over 10 realizations (median and 68\% credibility intervals) as a function of the number of episodes for \textit{(Top vs. Shirt)}-Fashion-MNIST. Since this is a much harder classification problem, the combined QKS+TTN protocol reaches an apparent noise floor at a much smaller qubit utilization. (b) Enforcing translational symmetry constraints in the TTN degrades performance in smaller architectures but offered equivalent or better performance in the largest architecture we studied.}
  \label{fig:figS3}
\end{figure}

\newpage
\subsection{Benchmarks on MNIST and Fashion-MNIST\label{app:multi-class}}
\begin{figure}[h!]
  \begin{center}
    \includegraphics[scale=1.5]{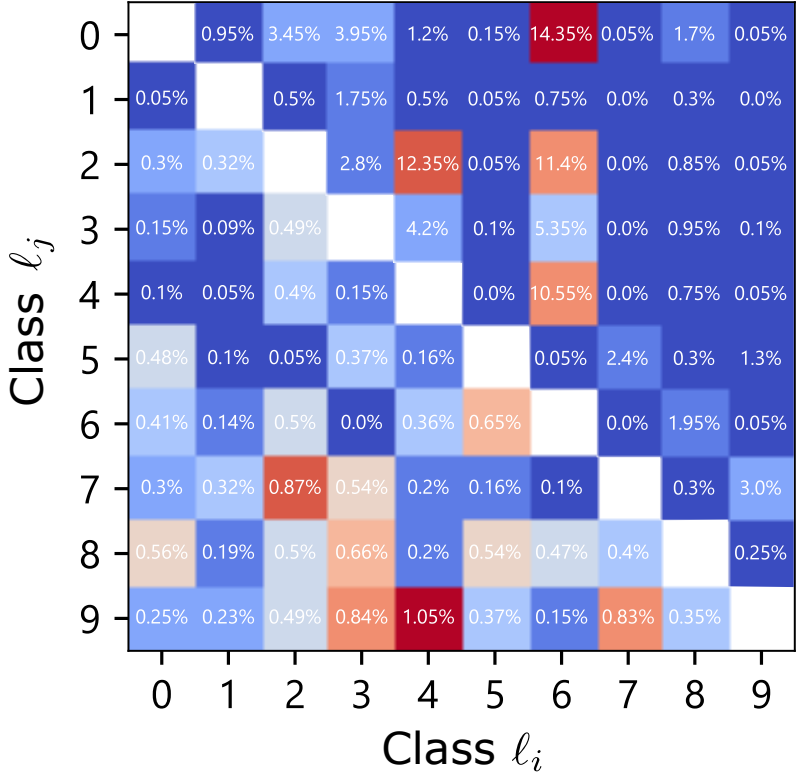}
  \end{center}
  \caption{Pairwise binary classification error rates for the best architectural configuration we tested---QKS+TTN+FO with translational symmetry, $E=512$, and $\chi=4$ trained for 40 epochs---on MNIST (lower triangle) and Fashion-MNIST (upper triangle).}
  \label{fig:figS4}
\end{figure}

As shown in the main text, the non-local feature maps of QKS led to significantly lower test errors than local feature maps on the hardest MNIST binary classification tasks. Here we have benchmarked the best architectural configuration we tested---QKS+TTN+FO with translational symmetry, $E=512$, and $\chi=4$ trained for 40 epochs---on the full MNIST and Fashion-MNIST datasets (Fig.~\ref{fig:figS4}). To the best of our knowledge, the model achieves state-of-the-art error rates compared to other variational quantum circuit and TN approaches~\cite{Huggins2019,Farhi2018,Chen2021,Hur2022}.

To test the generalization of QKS+TTN+FO architecture, we modified it directly for the multi-class case by mapping the full 16-dimensional output at the root node to 10 classes. At 512 episodes (no symmetry), the test error reached 1.82\%, while admitting a $45\times$ reduction in the number of parameters compared to the OvO construction; at 1,024 episodes, the test error reached $1.67\%$.

Since we described multi-class classification benchmarks for MNIST in the main text and above, here we focus on the multi-class classification of Fashion-MNIST. We established a baseline with QKS+SVM by constructing a multi-class model composed of one-versus-one (OvO) binary classifiers: at 10,000 episodes, the multi-class test error was 12.08\%. In comparison, QKS+TTN+FO with translational symmetry, trained as OvO binary classifiers, yielded a multi-class test error of 11.91\% at 512 episodes. 

\end{document}